\documentclass[a4paper,11pt]{article}
\usepackage{pos}
\usepackage{bm}\allowdisplaybreaks[1]
\usepackage{graphicx, xcolor}
\usepackage{adjustbox}
\usepackage{tikz}
\usetikzlibrary{arrows.meta}

\newcommand{\fig}[1]{Fig.~\ref{#1}}
\newcommand{\eq}[1]{Eq.~\eqref{#1}}

\newcommand{\pp}[1]{\left(#1\right)}
\newcommand{\bb}[1]{\left[#1\right]}

\newcommand{\beq}[1][]{\begin{equation}\label{#1}}
\newcommand{\eeq}{\end{equation}}
\newcommand{\bse}[1][]{\begin{subequations}\label{#1}}
\newcommand{\ese}{\end{subequations}}

\newcommand{\sgn}[1]{{\rm sgn}\left[#1\right]}
\newcommand{\wt}[1]{\widetilde{#1}}

\def\M{\mathcal{M}}

\title{New physical processes for extracting GPDs with a better sensitivity to partonic structure}

\author*[a,b]{Jian-Wei Qiu}
\author[a]{Zhite Yu}

\affiliation[a]{Theory Center, Jefferson Lab, Newport News, Virginia 23606, USA}

\affiliation[b]{Department of Physics, William \& Mary, Williamsburg, Virginia 23187, USA}

\emailAdd{jqiu@jlab.org}
\emailAdd{yuzhite@jlab.org}

\abstract{
We introduce a new type of exclusive processes for a better study of generalized parton distributions (GPDs), which we refer to as single-diffractive hard exclusive processes (SDHEPs). We advocate a two-stage framework for picturing SDHEPs based on the separation of scales, which gives a clear description both kinematically and dynamically. We examine the sensitivity of the SDHEP to the parton momentum fraction $x$-dependence of GPDs, and demonstrate it quantitatively with two specific processes that can be readily measured at J-PARC or AMBER using a pion beam and at JLab using a photon beam, respectively. Both processes are capable of providing enhanced sensitivity to the $x$-dependence, overcoming the problem of shadow GPDs, and disentangling different types of GPDs with various spin asymmetries.
}

\FullConference{31st International Workshop on Deep Inelastic Scattering (DIS2024)\\
 8–12 April 2024\\
Grenoble, France\\}

\begin{document}

\maketitle

\section{Introduction}
\label{sec:intro}
The research to explore internal structure of hadrons is entering a tomographic era 
owing to high-energy and high-luminosity accelerators together with theoretical and experimental advances to study
physical observables that have two distinct scales of momentum transfer in high energy collisions. 
With one hard scale $Q$ to localize the interaction with partons inside a hadron, a controllable soft scale 
can measure the motion and/or spatial distributions of quarks/gluons inside the probed hadrons.
In particular, by diffracting a colliding hadron from momentum $p$ to $p'$, hard exclusive processes with 
a soft scale $t = (p - p') \ll Q^2$ allow to extract the generalized parton distributions (GPDs)~\cite{Muller:1994ses, Ji:1996ek, Radyushkin:1997ki, Goeke:2001tz, Diehl:2003ny}, 
$F(x, \xi, t)$, with the parton's longitudinal momentum fraction $x$ and skewness $\xi = [(p - p') \cdot n] / [(p + p') \cdot n]$,
where $n$ is an auxiliary lightlike vector chosen to be conjugated to the momenta $p$ and $p'$~\cite{Qiu:2022pla}.
The Fourier transform of the GPDs' dependence on the soft scale $t$ at $\xi\to 0$ limit reveals the images of partons inside a confined hadron
$f(x, \bm{b}_T)$ at transverse spatial positions $b_T$ in slices of different $x$~\cite{Burkardt:2000za, Burkardt:2002hr}. 
The $x$ moments of GPDs $F_n(\xi, t) = \int_{-1}^1 dx x^{n-1} F(x, \xi, t)$ connect to various emergent properties of hadrons,
including their mass~\cite{Ji:1994av, Ji:1995sv, Lorce:2017xzd, Metz:2020vxd} and spin~\cite{Ji:1996ek} composition, 
internal pressure and shear force~\cite{Polyakov:2002yz, Polyakov:2018zvc, Burkert:2018bqq}.

However, extracting GPDs, especially their $x$ dependence, from experimental data is rather difficult~\cite{Ji:1996nm, Radyushkin:1997ki, Brodsky:1994kf, Frankfurt:1995jw, Berger:2001xd, Guidal:2002kt, Belitsky:2002tf, Belitsky:2003fj, Kumano:2009he, Kumano:2009ky, ElBeiyad:2010pji, Pedrak:2017cpp, Pedrak:2020mfm, Siddikov:2022bku, Siddikov:2023qbd, Boussarie:2016qop, Duplancic:2018bum, Duplancic:2022ffo, Duplancic:2023kwe, Qiu:2022bpq, Qiu:2023mrm, Qiu:2024mny}, 
suffering not only experimental but also theoretical challenges.
The difficulty is caused by the fact that GPDs of a diffracted hadron are defined at the amplitude level and 
the $x$-dependence represents the relative momentum of its active partons, integrated from $-1$ to $1$, 
while the exclusive scattering processes are more likely sensitive to $\xi$, proportional to the total momentum of the active partons.
This makes the extraction of GPDs a difficult inversion problem. 
In the following, we will discuss our proposal to overcome this intrinsic difficulty in extraction of GPDs.

\section{Single-diffractive hard exclusive processes}
\label{sec:sdhep}
\begin{figure}[htbp]
	\centering
	\begin{tabular}{cc}
		\includegraphics[trim={0 -2em 0 0}, clip, scale=0.6]{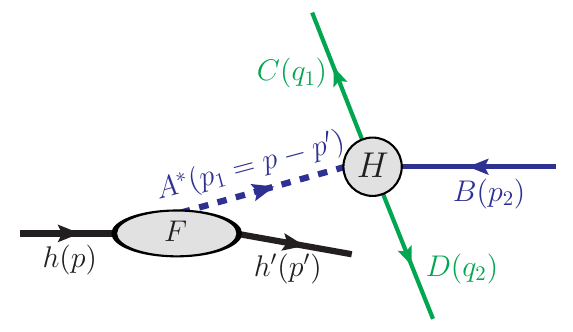} &
		\includegraphics[scale=0.45]{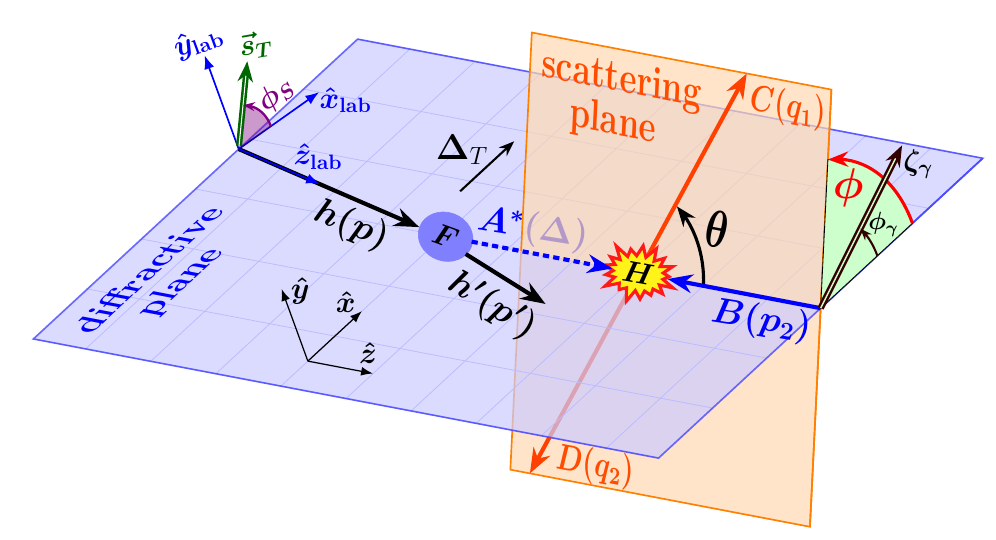} \\
		(a) & (b)
	\end{tabular}
	\caption{(a) Dynamic separation of the SDHEP amplitude into two stages. 
	         (b) The frame for analyzing SDHEPs in \eq{eq:sdhep}, where
		the linear polarization $\zeta_{\gamma}$ along $\phi_{\gamma}$ applies only to photoproduction processes.
		}
	\label{fig:sdhep}
\end{figure}

Before delving into the $x$-dependence problem, we first provide an overall discussion of physical processes involving GPDs.
Noticing that they have a diffraction ($t$) and hard interaction ($Q$) at the same time, 
the minimal kinematic configuration requires a $2 \to 3$ exclusive process,
\beq[eq:sdhep]
	h(p) + B(p_2) \to h'(p') + C(q_1) + D(q_2),
\eeq
which we refer to as single-diffractive hard exclusive process (SDHEP)~\cite{Qiu:2022pla},
where $h'$ is the diffracted hadron from $h$, 
$B$ is the beam particle that collides with $h$ 
to produce two particles $C$ and $D$ in the final state with large balancing transverse momenta in the center-of-mass (c.m.)~frame,
$q_T \sim q_{1T} \sim q_{2T} \gg \sqrt{-t}$,
with $q_T$ as the hard scale $Q$. 
To make the scale separation more manifest, we can picture the SDHEP as happening in two stages, 
\bse\label{eq:two-stage}\begin{align}
	&h(p) \to A^*(\Delta = p - p') + h'(p'), 	\label{eq:diffractive}\\
    &\hspace{9ex} \begin{tikzpicture}
        \node[inner sep=0pt] (arrow) at (0, 0) {
            \tikz{\draw[->, >={Stealth}, double, double distance=1pt, line width=1pt] (0, 0.28) to [out=-90, in=180] (0.8, 0);}
        };
    \end{tikzpicture} \hspace{1ex}
    A^*(\Delta) + B(p_2) \to C(q_1) + D(q_2),  \label{eq:hard 2to2}
\end{align}\ese
where the diffraction (\ref{eq:diffractive}) and the exclusive $2\to 2$ hard collision (\ref{eq:hard 2to2})
are connected by a virtual state $A^*$, as shown in Fig.~\ref{fig:sdhep}(a).  
The virtual state $A^*$ is much more long-lived than the time scale of the hard collision ($\sim 1/q_T$) 
due to its large momentum and low invariant mass.

This two-stage separation can be done both kinematically and dynamically. 
As shown in \fig{fig:sdhep}(b), we describe the SDHEPs in the c.m.~frame of the $2\to2$ hard scattering in \eq{eq:hard 2to2},
with the $\hat{z}$ axis along the direction of $A^*$ and $\hat{y} \propto \bm{p}' \times \bm{p}$ perpendicular to the diffraction plane.
This SDHEP frame is transformed from the Lab frame (the c.m.~frame of the $hB$ collision) by a trivial longitudinal boost 
followed by a power suppressed transverse boost~\cite{Diehl:2003ny}.
In this frame, it is natural to choose the auxiliary vector $n$ to be along the direction of $B$. 
Then $\xi$ becomes also a variable to describe the kinematics of the SDHEPs, together with $t$, $\phi_S$, $\theta$, and $\phi$,
where $\theta$ and $\phi$ are the polar and azimuthal angles of $C$ in the SDHEP frame.

As shown in \fig{fig:sdhep}(a), 
the scattering amplitude $\M$ can be formally written as a product or convolution of the amplitudes of the two subprocesses, 
summing over all possible $A^*$ states,
\beq[eq:channels]
	\mathcal{M}_{hB\to h'CD} = \sum_{n = 1}^{\infty} \sum_f F_{h\to h'}^{f_n}(p, p') \otimes C_{f_n B \to CD},
\eeq
where we organize in terms of the number $n$ and flavor $f$ of the particle contents in $A^*$, 
$F_{h\to h'}^{f_n}(p, p')$ describes the $h\to h'$ transition, 
and $C_{f_n B \to CD}$ stands for the hard $2\to2$ scattering amplitude.
For the simplest case, $n = 1$, $A^*$ can only be a virtual photon $\gamma^*$, 
which probes the electromagnetic form factors ($F_1$ and $F_2)$ of $h$.
This plays the role of the Bethe-Heitler process for the deeply virtual Compton scattering~\cite{Ji:1996ek, Radyushkin:1997ki}.
At $n = 2$, we have two-parton states, $A^* = [q\bar{q}]$ or $[gg]$, 
that form a loop to connect the diffraction and hard scattering subdiagrams in \fig{fig:sdhep}. 
At the leading power of $\sqrt{-t} / q_T$, the major contribution at $n = 2$ come from the loop momentum integration 
when the two active partons propagate almost collinearly.
By the exclusiveness, one can show that any soft gluons connected to $B$, $C$, and/or $D$ are cancelled 
or suppressed by powers of $\sqrt{-t} / q_T$~\cite{Qiu:2022pla},
so that the hard $2\to 2$ hard subdiagram (the probe) is factorized from the remaining part into GPDs, 
\beq[eq:factorization]
	\mathcal{M}^{[n = 2]}_{hB\to h'CD} = \sum_{i} \int_{-1}^1 dx \, F_{hh'}^{i}(x, \xi, t) \, C_{i B \to CD}(x, \xi; \hat{s}, \theta, \phi)
		+ \mathcal{O}\pp{\sqrt{-t} / q_T},
\eeq
where $\hat{s} = (\Delta + p_2)^2$ is the c.m.~energy squared of the $2\to2$ hard scattering 
and the $\sum_i$ runs over both parton flavors and GPD types.
The contributions from $n \geq 3$ channels are further suppressed by powers of $\sqrt{-t} / q_T$.

\section{Sensitivity to $x$ dependence of GPDs}
\label{sec:x}
Evidently from \eq{eq:factorization}, sensitivity to the GPD $x$ can only come from the hard coefficient $C$, 
which depends on the GPD variables $(x, \xi)$ and hard $2\to2$ kinematic observables $(\hat{s}, \theta, \phi)$,
among which the $\hat{s} \simeq 2 \xi s / (1 + \xi)$ is fixed by $\xi$ and the $\phi$ is determined by the spin states of $A^*$ and $B$,
\beq
	C_{A^* B \to CD}(x, \xi; \hat{s}, \theta, \phi) = e^{i (\lambda_A - \lambda_B) \phi} \, \bar{C}_{A^* B \to CD}(x, \xi; \hat{s}, \theta),
\eeq
where $\bar{C}_{A^* B \to CD}(x, \xi; \hat{s}, \theta) = C_{A^* B \to CD}(x, \xi; \hat{s}, \theta, \phi = 0)$, 
and $\lambda_A$ and $\lambda_B$ refer to the helicities of $A^*$ and $B$, respectively.
Only the measurement of $\theta$, or equivalently $q_T = (\sqrt{\hat{s}}/2) \sin\theta$, 
could be a probe of the GPD's $x$-dependence. 
For most known processes, however, the $x$ and $\theta$ dependence 
in the leading-order (LO) hard coefficient $\bar{C}$ are simply factorized from each other,
\beq[eq:scaling]
	\bar{C}(x, \xi; \hat{s}, \theta) = \sum_j G_j(x, \xi) \, T_j(\hat{s}, \theta),
\eeq
where $G_j(x, \xi) \propto (x \pm \xi)^{-1}$, with different $i \epsilon$ prescriptions determined by specific processes.
\eq{eq:scaling} reflects the scaling phenomenon at LO,
which implies that the sensitivity to GPDs is reduced to the simple moment integral,
\beq[eq:moment]
	F_0(\xi, t) = P\int_{-1}^1 dx \frac{F(x, \xi, t)}{x \pm \xi},
\eeq
where $P$ refers to principle-value integration,
and diagonal values $F(\pm \xi, \xi, t)$.
Getting the full $x$-dependence of GPDs from such moment-type sensitivity is almost impossible,
since one can construct a family of infinite functions $S(x, \xi, t)$, 
called shadow GPDs~\cite{Bertone:2021yyz, Moffat:2023svr},
satisfying $S_0(\xi, t) = 0$ and $S(\pm \xi, \xi, t) = 0$.
These shadow GPDs therefore cannot be separated from real GPDs by such processes.

To improve the sensitivity to the $x$-dependence of GPDs, we need 
to seek for processes that break the LO scaling in \eq{eq:scaling}. 
One well-known example is the double DVCS~\cite{Guidal:2002kt, Belitsky:2002tf, Belitsky:2003fj}, 
which unfortunately requires a very high integrated luminosity~\cite{Deja:2023ahc}. 
Here we propose two new processes that carry enhanced sensitivity.
One is the diphoton mesoproduction in nucleon-pion scattering~\cite{Qiu:2022bpq, Qiu:2024mny},
$	N(p) + \pi(p_2) \to N'(p') + \gamma(q_1) + \gamma(q_2)$,
and the other is the photon-pion pair photoproduction in nucleon-photon collision~\cite{Duplancic:2018bum, Duplancic:2022ffo, Duplancic:2023kwe, Qiu:2022pla, Qiu:2023mrm},
$	N(p) + \gamma(p_2) \to N'(p') + \pi(q_1) + \gamma(q_2)$.
In the hard scattering diagrams of the first process, the two photons can be radiated from different quark lines.
When $q_T\gg \sqrt{-t}$, these two quark lines must be connected by a hard gluon propagator that transmits the $q_T$ flow.
This generates a new GPD integral in addition to \eq{eq:moment},
\begin{align}
	I(\xi, t; z, \theta)
	= \int_{-1}^{1} dx \frac{F(x, \xi, t)}{x - x_p(\xi, z, \theta) + i \epsilon \, \sgn{\cos^2(\theta/2) - z} },
\label{eq:diphoton-special-int}
\end{align}
with a new pole $x_p$ that depends on $\theta$ (or equivalently, $q_T$),
\beq[eq:x-pole]
	x_p(\theta)
		= \xi \bb{ \frac{1 - z + \tan^2(\theta / 2) \, z}{1 - z - \tan^2(\theta / 2) \, z} }.
\eeq
Varying $\theta$ shifts this pole around in the DGLAP region of the GPDs, 
causing an entanglement between the GPD $x$ dependence and the observable $\theta$ (or $q_T$) distribution.
The second process is related to the first one by a kinematic crossing, so contains a similar integral to \eq{eq:diphoton-special-int}, 
but with the pole moving in the ERBL region, so gives a complementary sensitivity.
	
\begin{figure}[htbp]
	\centering
	\begin{tabular}{cc}
		\includegraphics[clip, scale=0.46]{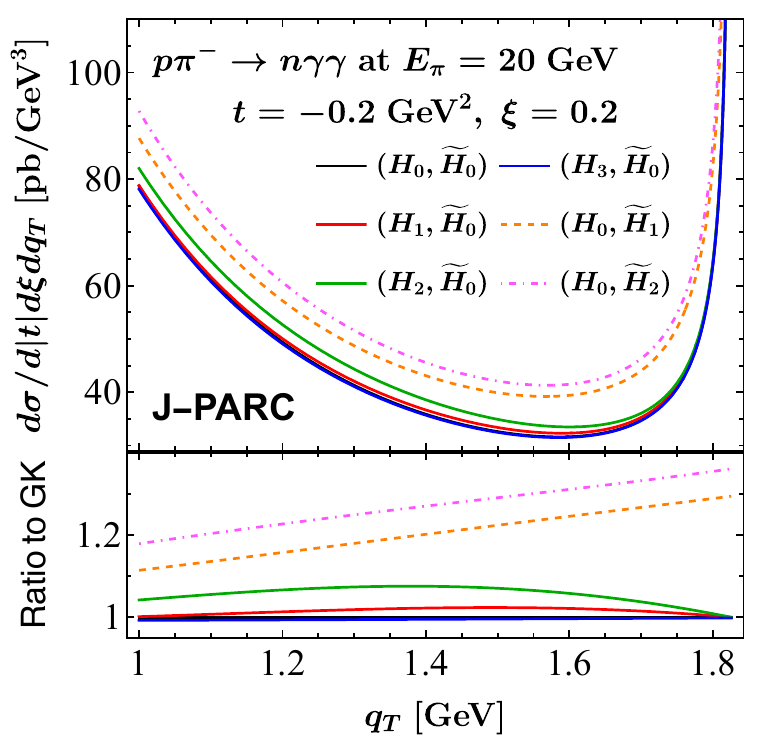} &
		\includegraphics[clip, scale=0.52]{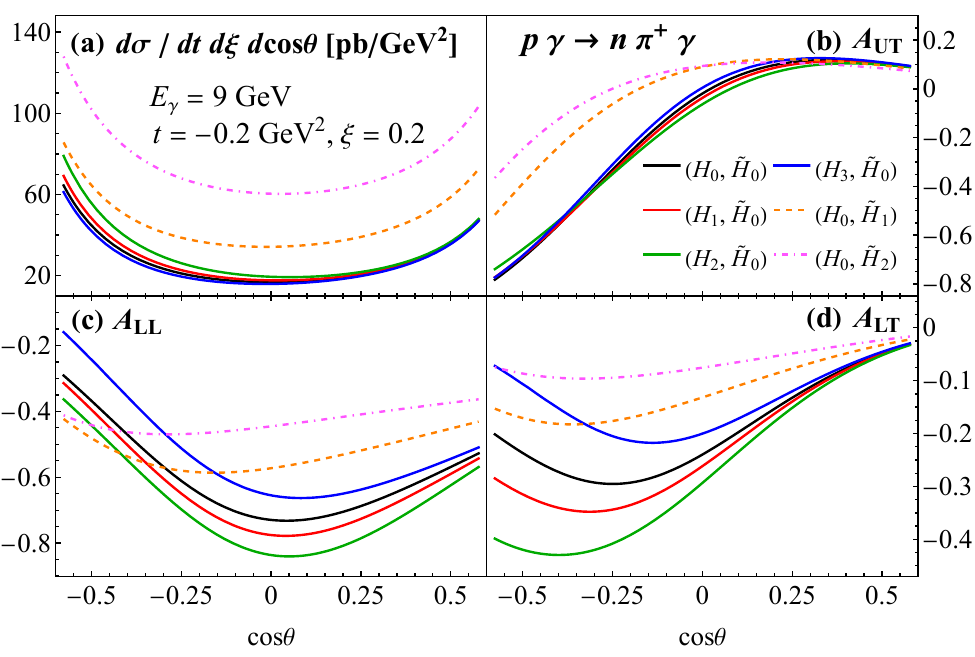} \\
		(a) & (b)
	\end{tabular}
	\caption{(a) The $q_T$ distributions for the diphoton cross section, and 
		(b) the $\cos\theta$ distributions for the photoproduction cross section and polarization asymmetries,
		evaluated using different GPD models.
	}
	\label{fig:qt-dist}
\end{figure}

Both processes can distinguish shadow GPDs from real ones. To demonstrate this, 
we construct some GPD models by starting with the Goloskokov-Kroll model~\cite{Goloskokov:2005sd, Goloskokov:2007nt, Goloskokov:2009ia, Kroll:2012sm} 
$(H_0, \wt{H}_0)$.
Adding analytically constructed shadow GPDs leads to a set of model GPDs
$(H_0, H_1, H_2, H_3)$ and $(\wt{H}_0, \wt{H}_1, \wt{H}_2)$.
We calculate the differential observables for both processes using these model GPDs.
For the diphoton mesoproduction, since pion is spin zero, one can only measure an unpolarized cross section (unless the proton carries a transverse spin).
This is shown in \fig{fig:qt-dist}(a) for the J-PARC kinematics.
For the photon-pion photoproduction, one can have additional polarization asymmetries when the photon is circularly or linearly polarized, 
giving more independent constraints on the GPDs, as shown in \fig{fig:qt-dist}(b) for the JLab Hall-D kinematics.
Clearly, in both cases, those different GPD models can be separated by measuring both the magnitude and shapes of the different observables.

\section{Conclusion}
\label{sec:conclusion}
We have proposed a unified description for GPD-related processes in terms of two-scale 
single-diffractive hard exclusive processes.
We also discussed how one should approach the $x$-dependence problem, and proposed 
two new processes that provide enhanced sensitivity to the GPD $x$-dependence.
Since there are eight different GPDs and each of them is a complicated function of three variables and of different parton flavors,
having two new processes is still not sufficient to map out all GPDs.  Looking forward,
it will be important to identify more physical processes, especially those with enhanced $x$-sensitivity,
and to incorporate with the lattice QCD data in a global analysis.
Awaiting us are both challenges and excitement.

\vspace{3mm}
This work is supported in part by the U.S. Department of Energy (DOE) Contract No.~DE-AC05-06OR23177, 
under which Jefferson Science Associates, LLC operates Jefferson Lab.

\bibliographystyle{apsrev}
\bibliography{reference}

\end{document}